\documentclass[pra,aps,longbibliography,showpacs,groupedaddress,superscriptaddress,twocolumn,toc=flat,nofootinbib]{revtex4-1}
\usepackage{graphicx}
\usepackage{latexsym}
\usepackage{amssymb}
\usepackage{amsmath}
\usepackage{amsfonts}
\usepackage{upgreek}
\usepackage{subfigure}
\usepackage{bm}
\usepackage{bbold}
\usepackage{verbatim}
\usepackage{siunitx}
\usepackage[unicode=true,
 bookmarks=false,
 breaklinks=false,pdfborder={0 0 1},backref=false,colorlinks=true]
 {hyperref}
\hypersetup{
 linkcolor=[rgb]{0,0,1},citecolor=[rgb]{0,0,1},urlcolor=[rgb]{0,0,1}}
\usepackage{multirow}
\usepackage{color}
\usepackage{comment}
\usepackage{xcolor}
\usepackage{soul}
\usepackage{tikz}
\usepackage{tabularx}
\usepackage{txfonts}
\allowdisplaybreaks

\renewcommand{\ij}{\langle i,j \rangle}

\newcommand{\Ztwo}{$\mathbb{Z}_2$}

\usepackage{braket}

\begin{document}

\newcommand{\mytitle}{Prethermal gauge structure and surface growth in \Ztwo{}~lattice gauge theories}
\title{\mytitle}

\author{Lukas~Homeier}
\email{lukas.homeier@jila.colorado.edu}
\affiliation{JILA, National Institute of Standards and Technology, and Department of Physics, University of Colorado, Boulder, CO, 80309, USA}
\affiliation{Center for Theory of Quantum Matter, University of Colorado, Boulder, CO, 80309, USA}

\author{Andrea~Pizzi}
\affiliation{Cavendish Laboratory, University of Cambridge, Cambridge CB3 0HE, United Kingdom}

\author{Hongzheng~Zhao}
\affiliation{State Key Laboratory of Artificial Microstructure and Mesoscopic Physics, School of Physics, Peking University, Beijing 100871, China}

\author{Jad~C.~Halimeh}
\affiliation{Department of Physics and Arnold Sommerfeld Center for Theoretical Physics (ASC), Ludwig Maximilian University of Munich, 80333 Munich, Germany}
\affiliation{Max Planck Institute of Quantum Optics, 85748 Garching, Germany}
\affiliation{Munich Center for Quantum Science and Technology (MCQST), 80799 Munich, Germany}
\affiliation{Department of Physics, College of Science, Kyung Hee University, Seoul 02447, Republic of Korea}

\author{Fabian~Grusdt}
\affiliation{Department of Physics and Arnold Sommerfeld Center for Theoretical Physics (ASC), Ludwig Maximilian University of Munich, 80333 Munich, Germany}
\affiliation{Munich Center for Quantum Science and Technology (MCQST), 80799 Munich, Germany}

\author{Ana~Maria~Rey}
\email{arey@jila.colorado.edu}
\affiliation{JILA, National Institute of Standards and Technology, and Department of Physics, University of Colorado, Boulder, CO, 80309, USA}
\affiliation{Center for Theory of Quantum Matter, University of Colorado, Boulder, CO, 80309, USA}

\date{\today}
\begin{abstract}
Universal aspects of thermalization in interacting many-body systems are challenging to derive microscopically, especially in kinetically constrained models, yet their numerical study beyond $(1+1)$D remains notoriously difficult. Here, we numerically study the mean-field dynamics of a $(2+1)$D spin system with thousands of spins and show that experimentally-feasible two-body Ising interactions can stabilize a prethermal \Ztwo{}~lattice gauge structure with dynamical matter, manifested by a separation of timescales with a stable gauge-invariant plateau. Eventually, the metastable prethermal \Ztwo{}~gauge structure breaks down via a proliferation of Gauss' law defects, similar to bubble formation in false vacuum decay. In this regime, we discover spatio-temporal correlations described by a non-linear surface growth consistent with the $(1+1)$D Kardar-Parisi-Zhang (KPZ) universality class, revealing a previously hidden feature in the thermalization of multi-point correlators. We benchmark our results in small systems against semi-classical discrete time Wigner approximation (DTWA) and exact diagonalization (ED), where the breakdown of DTWA signals the emergence of an extensive number of local symmetries that strongly influence the thermalization pathway.
Our model provides a testbed for quantum simulators and is directly implementable in large-scale arrays of Rydberg atoms.
\end{abstract}
\maketitle

\textit{Introduction.---}
How isolated many-body systems reach thermal equilibrium remains a central challenge in non-equilibrium physics~\cite{Nandkishore2015,Eisert2015,D’Alessio2016}, one that has recently become experimentally accessible through highly controlled quantum simulators~\cite{Kinoshita2006,Bernien2017,Choi2016,Wei2022,Zenesini2024,Andersen2025,Haghshenas2025}. These experiments allow one to probe universal properties of the dynamics and classify far-from-equilibrium phases of matter through, for example, hydrodynamic descriptions~\cite{Pruefer2018,Erne2018,Johnstone2019,Navon2019,Glidden2021,GarciaOrozco2022,Huh2024,Martirosyan2024}.

In recent years, lattice gauge theories (LGTs), originally developed in the context of high-energy physics, have become a prominent playground to explore non-trivial dynamics, due to their enriched structure rooted in local constraints~\cite{Banuls2020,Halimeh2025,Halimeh2025_OutOfEq,Dalmonte2016}, with prominent examples observed in far-from-equilibrium experiments~\cite{Bernien2017,Surace2020,Su2023a,Smith2017,Brenes2018,Gyawali2025,Smith2017,Brenes2018,Gyawali2025,Martinez2016,GonzalezCuadra2025,Cochran2025,De2024}.
This makes LGTs a versatile framework to study non-equilibrium phenomena, some of which have also been observed in a broader class of strongly-interacting many-body systems, from Hubbard models~\cite{Schreiber2015,Adler2024} to spin systems~\cite{Bernien2017,Ljubotina2019,Wei2022} and superfluids~\cite{Zenesini2024,Zhu2024}.

Despite these challenges, a universal description of thermalization in LGTs is largely unexplored -- particularly for kinetically constrained models~\cite{Gromov2020,Chakraborty2025}.
As the field matures, significant effort has been devoted to find physically relevant phenomena accessible in simple experimental implementations of LGTs~\cite{Halimeh2025_OutOfEq}. 
However, theoretical guidance and concrete predictions are notoriously difficult, since numerical simulations of $(2+1)$~LGTs are currently limited to short times or small system sizes~\cite{Banuls2023}. Overcoming these limitations requires new approximate approaches that provide access to large system sizes and long times, and that can be benchmarked and tested in quantum simulation experiments.


Here, we employ efficient classical spin simulations of a \Ztwo{}~LGT with dynamical matter over large space and time scales to unveil the universal features of its thermalization. The model is motivated by its experimental feasibility in Rydberg arrays~\cite{Homeier2023}, and constructed such that the gauge-invariant subspace arises as a metastable manifold, protected by an energy gap~$\propto V$. In the presence of a weak gauge-symmetry-breaking drive~$\Omega$, the system exhibits a long prethermal gauge-invariant plateau, which emerges above a critical protection strength $V_c$, relevant to the large-scale quantum simulation of gauge theories~\cite{Halimeh2025}. At late times, thermalization consists of the breakdown of gauge symmetry, which we show to be governed by the nucleation and growth of gauge defects. A careful scaling analysis of the surface growth shows critical exponents consistent  with the $(1+1)$D KPZ universality class~\cite{Kardar1986,CORWIN2012}; a hidden feature in the spreading of multi-point correlation functions. We further benchmark small systems beyond mean-field dynamics using DTWA~\cite{Schachenmayer2015} and ED. Surprisingly, we find that, while the mean-field dynamics shows qualitative agreement with ED, DTWA is not able to capture the prethermal plateau, highlighting how, unlike in common cases, thermalization in LGTs cannot be understood as a simple phenomenon of scrambling of fluctuations, but is instead strongly affected by the local (emergent) dynamical constraints.

\begin{figure}[t!!]
\centering
\includegraphics[width=\linewidth]{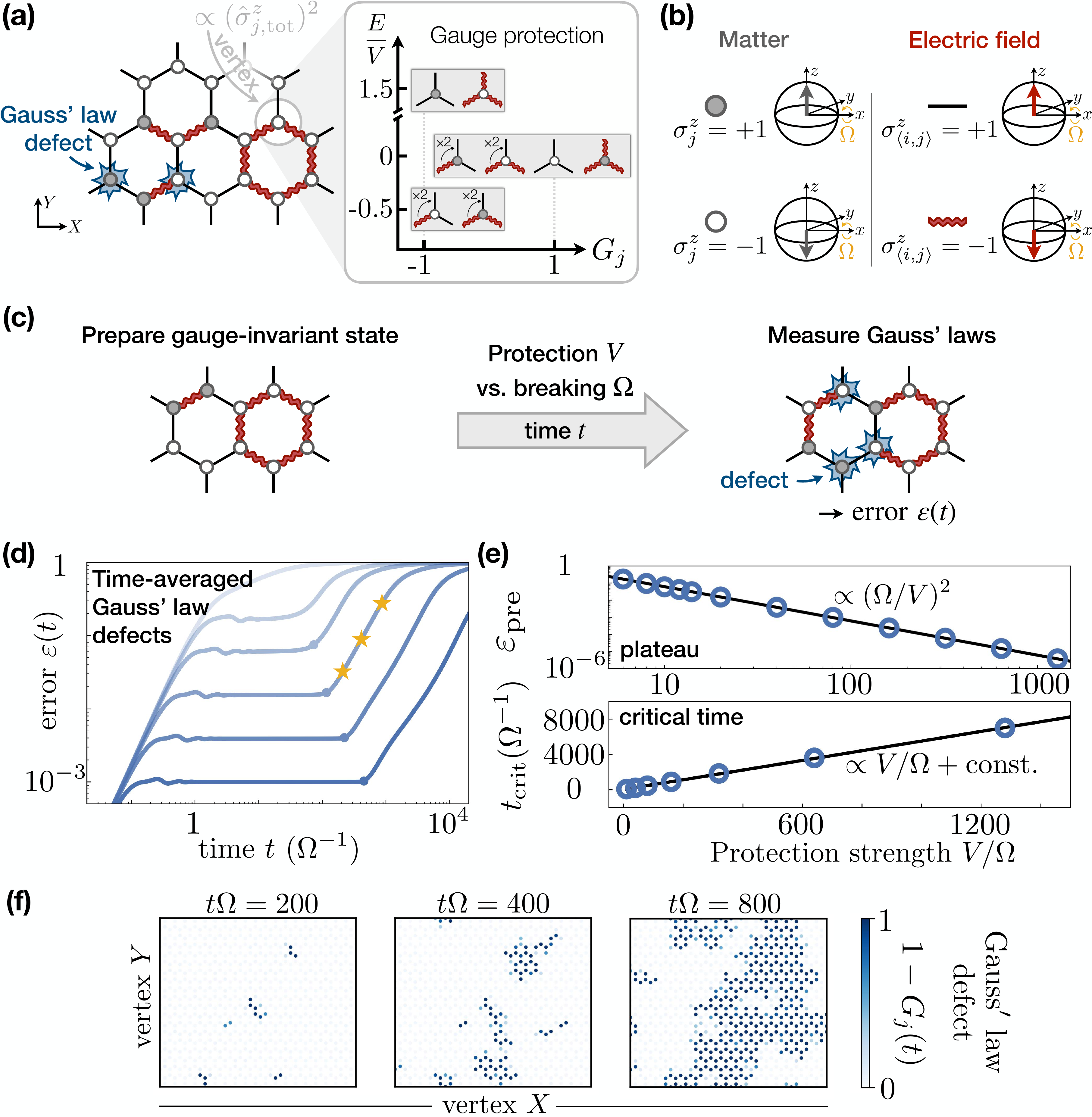}
\caption{\textbf{Spin model and emergent gauge structure.} \textbf{(a)} Spins are located on the sites (matter) and links (electric/gauge fields) of a honeycomb lattice. We define a \Ztwo{}~Gauss' law, see Eq.~\eqref{eq:Gauss-law}, together with a two-body Hamiltonian~$\hat{H}_V$ that energetically protects a metastable sector at energy~$E=0$ containing the gauge-invariant configurations~$G_j=+1$ (inset). \textbf{(b)}, \textbf{(c)} The mean-field dynamics describe the evolution of classical spins, i.e., unit vectors on a sphere. After the system is prepared in a random gauge-invariant state, the perturbation~$\hat{H}_\Omega$ rotates the spins about the $x$ axis, leading to the dynamical violation of the local Gauss' law constraints at later times~$t$. \textbf{(d)} We evolve gauge-invariant initial states for various protections strengths $V/\Omega=4,6,10,20,40,80$ (light to dark color) averaging over $200$ random realizations in a $20\times 20$ plaquette system. The gauge-symmetry breaking errors remain small and controlled for a long time, leading to prethermal plateau with an emergent \Ztwo{}~gauge structure. The critical time~$t_{\rm crit}$ of the plateau is defined by the maximum curvature (blue dots). \textbf{(e)} The protection strength~$V/\Omega$ controls the magnitude (top) and critical time (bottom) of the prethermal plateau; black lines indicate fit functions. \textbf{(f)} Local Gauss' laws of one exemplary trajectory for multiple times after the plateau phase, see yellow markers in panel \textbf{(d)}. Local nucleation regions followed by a proliferation of Gauss' law defects imply a rich spatio-temporal correlation structure. }
\label{figure-1}
\end{figure}

\textit{Model.---}We consider spins~$\hat{\bm{\sigma}} = (\hat{\sigma}^x,\hat{\sigma}^y,\hat{\sigma}^z)$ on the sites~$j=(X,Y)$ and links~$\ij$ of a honeycomb lattice, see Fig.~\ref{figure-1}(a,b). By analogy with Ising LGTs~\cite{Wegner1971,FradkinSusskind1978,FradkinShenker1979}, the matter sites can be occupied by \Ztwo{}~charges ($\hat{\sigma}^z_j=+1$) and the links host \Ztwo{}~electric fields ($\hat{\sigma}^z_{\ij}=-1$), for which we define the \Ztwo{}~Gauss' law on vertex~$j$
\begin{align} \label{eq:Gauss-law}
    \hat{G}_j = -\hat{\sigma}^z_j \prod_{i: \ij} \hat{\sigma}^z_{\ij}
\end{align}
containing the matter site~$j$ and its adjacent links, such that $G_j:=\langle \hat{G}_j \rangle \in [-1,1]$; a vertex~$j$ is fully gauge symmetric (symmetry breaking) if~$G_j = +1$ ($G_j=-1$).
A Hamiltonian has a local \Ztwo{}~symmetry if~$[\hat{H}, \hat{G}_j ]=0$ for all~$j$, such that the Gauss' laws are a constant of motion,~$\dot{G_j}=0$.

Motivated by its experimental relevance~\cite{Homeier2023}, we consider a Hamiltonian $\hat{H} = \hat{H}_{\mathbb{Z}_2} + \hat{H}_V + \hat{H}_\Omega$, with
\begin{align}
\begin{split} \label{eq:Hamiltonian}
    \hat{H}_{\mathbb{Z}_2} &= J\sum_{\ij} \hat{\sigma}^x_i \hat{\sigma}^{x}_{\ij}\hat{\sigma}^x_j, \\
    \hat{H}_V &= V\sum_j \frac{(1+\delta_j)}{8} \left[\left( \hat{\sigma}^z_j + \sum_{i:\ij}\hat{\sigma}^z_{\ij} \right)^2 - 4 \right], \\
    \hat{H}_\Omega &= \frac{\Omega}{2}\left( \sum_j \hat{\sigma}^x_j +  \sum_{\ij} \hat{\sigma}^x_{\ij} \right).
\end{split}
\end{align}
The first term~$\hat{H}_{\mathbb{Z}_2}$ is a gauge-invariant coupling between matter and gauge fields, typical for Ising \Ztwo{}~LGTs~\cite{FradkinSusskind1978,FradkinShenker1979}.
The second term $\hat{H}_V$ describes a local \textit{pseudo}generator~\cite{Homeier2023,Halimeh2022LPG}, composed of two-body interactions in the $z$ basis. Up to an overall energy shift, this term is proportional to $\sum_j  (\hat{\sigma}^z_{j,\rm tot})^2$, where~$\hat{\sigma}^z_{j,\rm tot}$ is the total magnetization of all spins within vertex~$j$. For~$\delta_j=0$, the term engineers an energy landscape where the gauge symmetric vertices contribute no energy, and the gauge-symmetry breaking vertices contribute an energy~$-0.5V$ or $1.5V$, see Fig.~\ref{figure-1}(a). That is, $\hat{H}_V$ introduces a gauge protection via energy penalties~\cite{Halimeh2021PRXQ}. The third term~$\hat{H}_\Omega$ describes rotations about the $x$ axis, breaking the gauge symmetry, see Fig.~\ref{figure-1}(b).

In the following, we will mostly focus on the mean-field dynamics $\dot{\vec{\sigma}} = \{\vec{\sigma},H\}$, where $\vec{\sigma}$ denotes a classical unit vector satisfying Poisson brackets $\{\sigma^a, \sigma^b\} = \varepsilon_{abc} \sigma^c$, with $\varepsilon_{abc}$ the Levi-Civita symbol.
Conveniently, the classical mean-field dynamics can be efficiently solved for large $(2+1)$D~systems. In the last part, we will compare it to semi-classical DTWA and ED in small systems.

\textit{Prethermal \Ztwo{}~LGT.---}
Initializing the system in a configuration with~$G_j=+1$ for all~$j$, we expect the system to remain gauge-invariant for a long time if~$\Omega \ll V$. For~$\delta_j=0$, the breaking of the Gauss' law constraints will ultimately happen through correlated processes on four vertices due to resonant processes where one vertex increases energy by $+1.5V$ and three vertices decrease their energy by~$-0.5V$, see Fig.~\ref{figure-1}(a). The duration of the gauge invariance can be further extended lifting these degeneracies with random couplings~$\delta_j$, e.g., normally distributed around zero with standard deviation~${\delta = 0.1}$~\cite{Homeier2023}; see the End Matter for a discussion of various disorder strengths.

We probe the breakdown of the gauge symmetry by numerically integrating the mean-field dynamics on a~$20 \times 20$ plaquette system ($N=2000$ spins, $N_{\rm vert}=800$ vertices and $3N$ coupled differential equations) with periodic boundaries and using trotterization techniques~\cite{Howell2019,Pizzi2021}, see End Matter for details. We consider an ensemble of~$200$ trajectories, each with a different random gauge-invariant initial configuration and set of couplings~$\delta_j$. The gauge-symmetry breaking is quantified by the time-averaged error~$\varepsilon(t) = t^{-1} \int_{0}^t \mathrm{d}\tau [1- \langle G_j(\tau) \rangle]$, where~$\langle 
\cdot\rangle$ denotes the average over the ensemble and vertices, see Fig.~\ref{figure-1}(c).
Its behavior is shown in Fig.~\ref{figure-1}(d) for various protection strengths~$V/\Omega = 4, \ldots, 80$; in the following, we set~$J = \Omega$. 

We find three distinct regimes~\cite{Halimeh2020PRLRealiability}: (i) At early times~${t \Omega \lesssim 1}$, the error shows a universal growth $\varepsilon(t) \propto t^2$. (ii) The error settles to a stable plateau with small gauge-symmetry breaking for a time linear in~$V/\Omega$. Hence, the system establishes a prethermal phase with an emergent \Ztwo{}~gauge symmetry. In Fig.~\ref{figure-1}(e) we plot the value $\varepsilon_\mathrm{pre} := \varepsilon(t\Omega=20)$ of the prethermal plateau and find that it follows a power-law~$\varepsilon_\mathrm{pre} \propto (\Omega/V)^2$ controlled by the protection strength~$V$~\cite{Halimeh2022LPG}. 
The critical time after which the plateau becomes unstable is defined as the time of maximum curvature after the plateau, see blue dots in Fig.~\ref{figure-1}(d), and scales as~$t_\mathrm{crit} \propto V/\Omega$. (iii) At late times, the \Ztwo{}~Gauss' law breaks down and the system reaches the infinite temperature value, $\varepsilon \rightarrow 1$. Finally, an examplary trajectory in Fig.~\ref{figure-1}(f) shows Gauss' law violations with strong spatio-temporal correlations, which we will discuss in detail below.

The existence of a prethermal phase with small and controlled gauge-symmetry breaking errors is crucial for the gauge-invariant quantum simulation of \Ztwo~LGTs. Going beyond $(1+1)$D systems~\cite{VanDamme2025} and small-scale ED predictions~\cite{Homeier2023,Halimeh2022LPG,Halimeh2022_DFL}, our results show a promising method to stabilize gauge constraints in large-scale $(2+1)$D quantum simulators~\cite{Homeier2023}.


\begin{figure}[t!!]
\centering
\includegraphics[width=\linewidth]{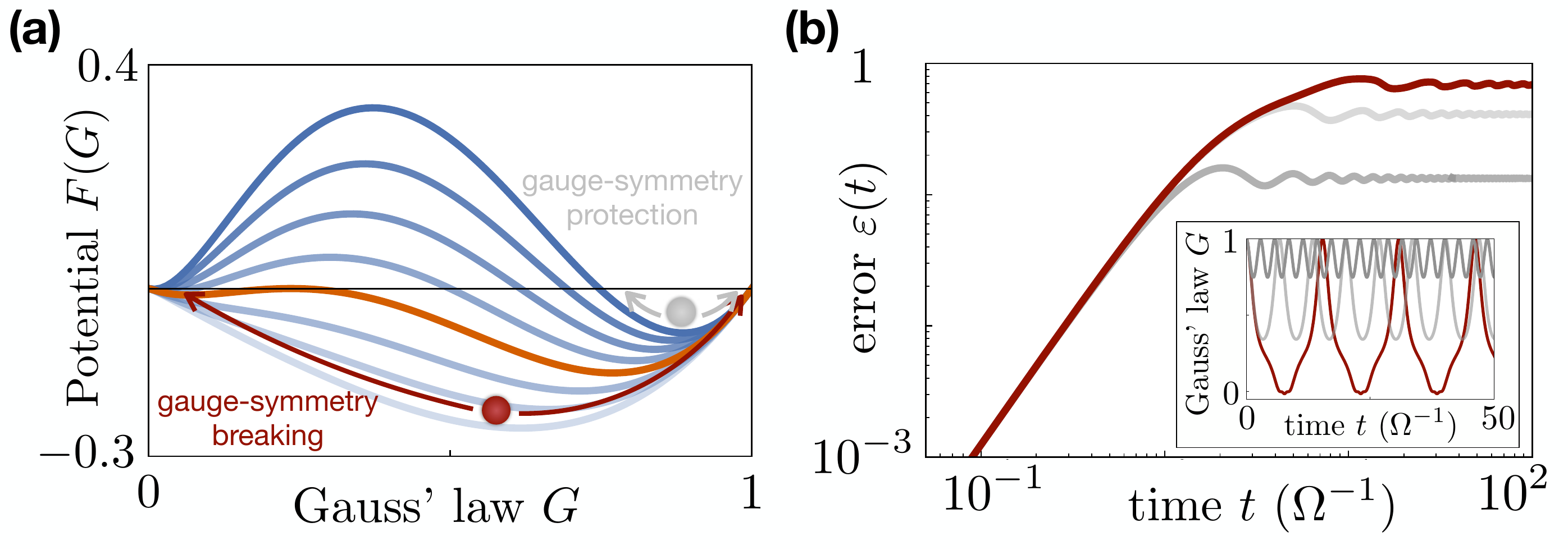}
\caption{\textbf{Mean-field phase transition.} \textbf{(a)} Potential~$F(G)$ for the translationally-invariant model without dynamical matter, for protection strengths~$V/\Omega = 2,..,5$ (bright to dark) and $(V/\Omega)_c \approx 3.33$ (orange). The dynamics is governed by the kinetic equation $(\dot{G})^2 + F(G) = 0$ with initial conditions~$G(0)=+1$ and $\dot{G}(0)=0$. \textbf{(b)} For a protection strength below the critical value, the error grows to its maximum in the translationally-invariant setting (red curve). As the protection is increased, the error is suppressed since the root in the potential $F(G)$ moves towards $G=1$. Inset: The Gauss' law shows large gauge-symmetry breaking oscillations for $(V/\Omega)=3.3 < (V/\Omega)_c$ (red curve), and constrained dynamics for $(V/\Omega)=3.4 > (V/\Omega)_c$ (light gray curve) with smaller oscillations for increasing protection, e.g., $V/\Omega=5$ (dark gray curve). }
\label{figure-2}
\end{figure}
\textit{Critical protection.---}
Next, we gain analytic insights and physical intuition into the prethermalization behavior by studying the simplest possible setting. This allows us to predict a phase transition, where above a critical protection strength~$(V/\Omega)_c$ the dynamics exhibits a prethermal plateau.
To this end, we consider the translationally-invariant model without dynamical matter ($\sigma^z_j(t) \equiv -1$ for all~$t$ and $J=0$), and we derive the mean-field equations of motion for the initial state with no electric field lines ($\sigma^z_{\ij}(0) = +1$).
Conservation of energy and spin length yields the kinematic equation for Gauss' law~$\dot{G}^2 + F(G) = 0$, where~$F(G) = F(G; \Omega, V)$ is a potential energy, see End Matter for a derivation. In Fig.~\ref{figure-2}(a), we plot the potential for various protection strengths~$V/\Omega$: Starting from a gauge-invariant state~$G=+1$, the gauge-breaking dynamics is unconstrained when~$V/\Omega$ is below a critical protection strength~$(V/\Omega)_c  = \sqrt{\frac{11+5\sqrt{5}}{2}}\approx 3.33$, and the Gauss' law shows strong oscillations, see Fig.~\ref{figure-2}(b). Above~$(V/\Omega)_c$, the dynamics becomes increasingly constrained and oscillates around a value close to~$G=+1$, which is associated with the development of an additional root in the potential~$F(G)$ shown by the orange curve in Fig.~\ref{figure-2}(a).

\begin{figure}[t!!]
\centering
\includegraphics[width=\linewidth]{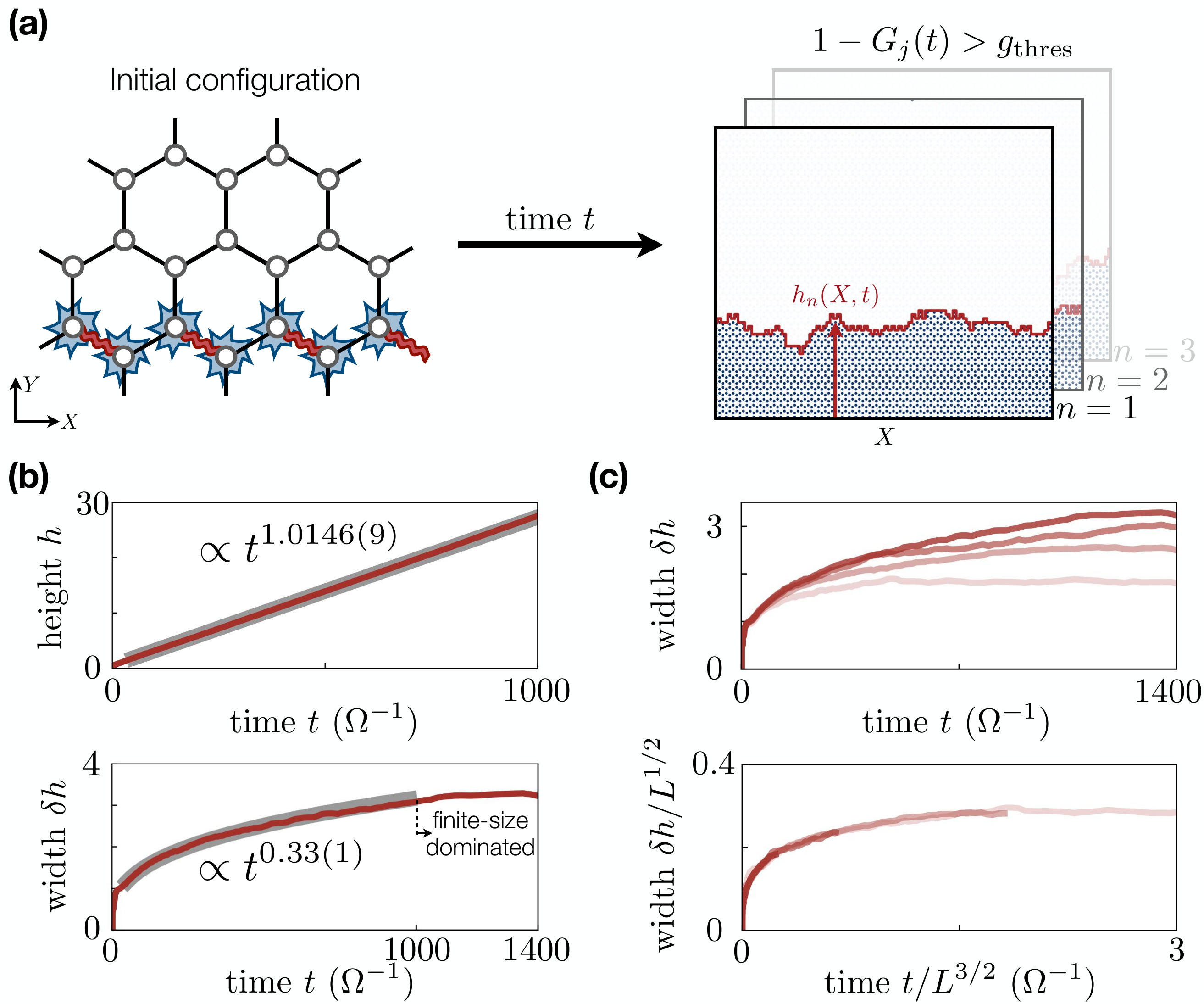}
\caption{\textbf{Surface growth.} \textbf{(a)} We investigate the dynamics starting from an initial configuration with Gauss' law defects in the bottom row. We find defect and defect-free regions separated by a surface (red line, right) that we quantify by the height function~$h_n(X,t)$, where $n$ labels a trajectory for a set of random protection strengths~$\{\delta_j \}$; here we set~$V/\Omega=10$. \textbf{(b)} We analyze the mean height~$h$ and width~$\delta h$, Eq.~\eqref{eq:height_function}, showing ballistic spreading and a non-linear roughening, respectively; statistical error bars of the numerical data are smaller than the solid lines. The gray lines show power-law fits to the data for $50 < t 
\Omega < 1000$, where the width~$\delta h$ does not yet crossover to the asymptotic Family-Vicsek behavior. \textbf{(c)} Top: Width~$\delta h$ for increasing linear system size~$L=40,80,160,320$ (light to dark). Bottom: The dynamical scaling analysis reveals an excellent collapse of the data for a roughening exponent~$\alpha=1/2$ and dynamical exponent~$z=3/2$.     }
\label{figure-3}
\end{figure}
\textit{Surface growth.---}The trajectory of an initial gauge-invariant configuration reveals a characteristic behavior, see Fig.~\ref{figure-1}(f): Upon exiting the prethermal plateau, random fluctuations, determined by the initial configuration and the couplings~$\delta_j$, lead to the nucleation of gauge-symmetry breaking errors. Subsequently, neighboring vertices violate the Gauss' law constraint, so that the errors proliferate until the entire system reaches equilibrium, suggesting a universal hydrodynamic description based on a surface growth model~\cite{Corwin2016}. The role of spatio-temporal correlations have been previously discussed in kinetically constrained models in $(1+1)$D~\cite{Michailidis2018,Causer2024,Maric2026}.

To test our hypothesis, we perform numerical simulation on systems with $50 \times 50$ plaquettes and periodic (open) boundaries in the $X$ ($Y$) spatial direction.
We initialize a configuration with a line defect in the bottom row, see Fig.~\ref{figure-3}(a) left, and repeat the simulations for~$N_{\rm Ens}=200$ realizations of random couplings~$\delta_j$~\cite{Footnote_boundary_term}.
To each space-time point we associate a binary variable $\Theta(1-G_j(t) -g_{\rm{thres}})$, where $\Theta(\cdot)$ is the Heaviside function; in the following we set~$g_{\rm{thres}}=0.2$ but our findings do not depend on this specific value.
This binary variable defines a surface separating the regions of small and large Gauss' law violations, see Fig.~\ref{figure-3}(a).
Further, we use the surface to define a height function $h_n(X,t)$ in Euclidean space~\cite{Footnote_euclidean} and for each realization~$n=1,\ldots,N_{\rm Ens}$.

Next, we analyze the height function and the cumulants associated with this distribution, such as the mean~$h(t)$ and width~$\delta h(t)$ given by
\begin{align}
\begin{split} \label{eq:height_function}
h(t) &= \big\langle \langle{h}_n(X,t)\rangle_X \big\rangle_n, \\
\delta h(t) &= \bigg\langle \sqrt{\big\langle\left[ h_n(X,t) - \langle h_n(X,t)\rangle_{X} \right]^2 \big\rangle_X} \bigg\rangle_n,
\end{split}
\end{align}
where~$\langle \cdot \rangle_n$ and~$\langle \cdot \rangle_X$ denote averaging over the ensemble and over the dimension $X$, respectively; for the case of $320$ plaquettes the linear dimension is $L=640$.

In Fig.~\ref{figure-3}(b), we fit power-law functions~$\propto a + b t^{c}$ to the mean height~$h(t)$ and width~$\delta h(t)$ for times $ 50 <t\Omega < 1000$ (grey curves). Such an intermediate time range limits the non-universal behavior at short times and finite-size effects at long times. We observe a ballistic spreading of the mean height~$h(t) \propto t^{1.0146(9)}$ and non-linear roughening of the surface width $\delta h(t) \propto t^{\,\beta}$ with growth exponent $\beta=0.33(1)$. The error bars describe one standard deviation of a least square minimization fitting procedure, and our simulations are likely dominated by systematics, such as finite size effects, more than by statistical fluctuations. To confirm that our model is consistent with ballistic deposition~\cite{Corwin2016} and KPZ scaling, we carefully take finite size effects into account performing a dynamical scaling analysis.

For this, we consider the Family-Vicsek scaling relation $\delta h \sim L^{\alpha} f(t/L^z)$, where $\alpha$ is the roughening exponent and $z=\alpha/\beta$ is the dynamic exponent~\cite{Family1985}. The dynamical scaling captures the spreading of spatial correlations, which saturate after a time $t \sim L^z$. In Fig.~\ref{figure-3}(c) top, we show the width~$\delta h$ for increasing system sizes~$L=40,80,160,320$. By rescaling with the Family-Vicsek relation, we obtain a collapse of the data using the exponents~$\alpha=1/2$ and~$\beta=1/3$ of the $(1+1)$D KPZ universality class~\cite{Kardar1986,Corwin2016}, see Fig.~\ref{figure-3}(c) bottom. 

We note that a complete characterization of the universality class governing the surface growth necessitates to evaluate higher-order cumulants~\cite{Ljubotina2019,Wei2022,Rosenberg2024}, and that in our simulations the skewness and kurtosis of the height field are not converged. Despite these challenges, that future studies should address considering larger systems, our findings strongly suggest KPZ universality of the spreading of Gauss' law defects. 
While spatio-temporal correlations remain hidden in multi-point correlation functions, they are naturally revealed within our gauge-theoretical interpretation of the transverse- and longitudinal-field Ising model on the honeycomb lattice.

An alternative interpretation to the proliferation of Gauss' law defects is provided by introducing a generalized Gauss' law~$\hat{\tilde{G}}_j$ under which all terms in the Hamiltonian become gauge invariant, i.e., $[\hat{H}, \hat{\tilde{G}}_j]=0$ and the term~$\hat{H}_\Omega$ corresponds to pair creation processes mediated by single-body operators. In this way, our results can be directly connected to false-vacuum decay~\cite{Zhu2024} and the thermalization dynamics of LGTs with dynamical matter~\cite{Zhou2022,Mueller2025}.

\begin{figure}[t!!]
\centering
\includegraphics[width=\linewidth]{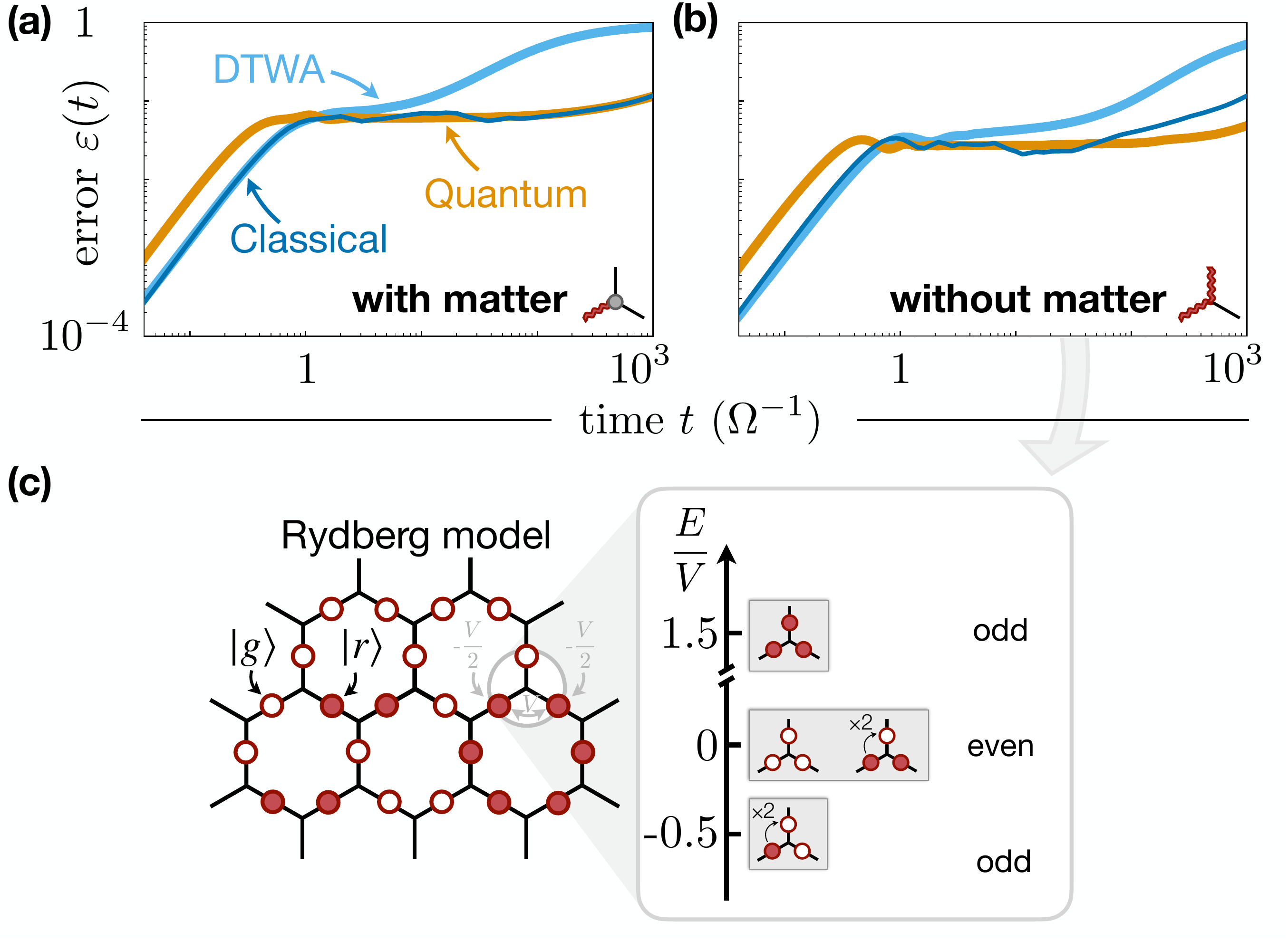}
\caption{\textbf{Quantum dynamics.}  \textbf{(a,b)} We compare the dynamics in small systems using classical mean-field dynamics, DTWA, and ED for models with \textbf{(a)} and without \textbf{(b)} dynamical matter, for~$J=0$ and~$\delta_j=0$. In both cases, the mean-field dynamics has excellent agreement with ED, whereas the DTWA only marginally captures the prethermal plateau due to the emergent local \Ztwo{}~symmetry. \textbf{(c)} The model without dynamical matter is naturally realized in Rydberg tweezer arrays with nearest-neighbor density-density interactions of strength~$V$ between excited state atoms~$\ket{r}$ and a laser drive~$\Omega$ between $\ket{g}$ and $\ket{r}$. By choosing the detuning of the drive in the facilitation-like regime, $\Delta=V$, an even number of excitations sector is stabilized. The energetic separation of even and odd excitation number sectors stabilizes the \Ztwo{}~symmetry. }
\label{figure-4}
\end{figure}
\textit{(Semi)classical and quantum model.---}The mean-field dynamics considered thus far is exact in the large-spin~$S \rightarrow \infty$ limit~\cite{Mori2018}. We expect that a similar KPZ universality in the thermalization dynamics should appear for a spin-$1/2$ \Ztwo{}~LGT. This could be probed experimentally in state-of-the-art Rydberg tweezer experiments~\cite{Browaeys2020} with hundreds or thousands of qubits~\cite{Manetsch2025,Gyger2024,Chiu2025}. Here, given the notorious numerical challenges of quantum many-body systems~\cite{Banuls2023}, we study the long-time dynamics in small systems of~$L=2 \times 2$ plaquettes with periodic boundary conditions, comparing mean-field dynamics, DTWA, and ED.

In particular, we focus on the experimentally relevant case of only two-body interactions~\cite{Footnote_three_body_term} ($J=0$), no disorder ($\delta_j=0$), and we set~$V/\Omega=10$. 
Analogous to above, we compute the time and ensemble averaged error~$\varepsilon(t)$ predicted by the three methods in a model with matter, see Fig.~\ref{figure-4}(a), and without matter (setting ${\hat{\sigma}^z_j\equiv 1}$ for all $j$), see Fig.~\ref{figure-4}(b).
While the system sizes are too small to analyze the surface growth, we find a prethermal \Ztwo{}-invariant plateau for both classical dynamics and quantum ED. The remarkable, qualitative agreement in our case is possible because the Abelian \Ztwo{}~Gauss' law is a non-connected correlator without genuine quantum correlations. 
Notably, the semi-classical DTWA fails to describe the quantum case and is outperformed by the mean-field dynamics. We attribute this behavior to the dynamical emergence of symmetry constraints imposed by the protection term~$\hat{H}_V$. A potential cause of the DTWA breakdown is that the added stochastic noise, which reproduces the full density matrix of the initial state~\cite{Schachenmayer2015}, can seed the occupation of gauge-symmetry breaking sectors~\cite{Muleady2023} due to the inappropriate evolution of the higher-order correlators. This  failure highlights that pre-thermalization in gauge theories  strongly depends on the subtle interplay between scrambling of fluctuations and locally constrained dynamics.

Lastly, we describe how the model studied in Fig.~\ref{figure-4}(b) can be implemented in quantum simulators--we highlight that this particular regime presents the simplest instance to realize the kinetic constraints. Since the matter spins on the sites are frozen, we remove them to obtain a model with spins only residing on the links of the honeycomb lattice, i.e., on the sites of the dual Kagome lattice, yielding the Hamiltonian
\begin{align} \label{eq:Rydberg_Ham}
    \hat{H}^{\rm Ryd} = \frac{V}{4}\sum_{\langle A,B \rangle} \hat{\sigma}^z_{A}\hat{\sigma}^z_{B} + \frac{V}{2}\sum_A \hat{\sigma}^z_{A} + \frac{\Omega}{2}\sum_A\hat{\sigma}^x_{A},
\end{align}
where $\langle A,B \rangle$ denote nearest-neighbour sites on the Kagome lattice.
The model is readily implementable in Rydberg tweezer arrays~\cite{Osterholz2025,Chao2025} by identifying~$\hat{\sigma}^z_{A}=-1$ ($\hat{\sigma}^z_{A}=+1$) with the atomic ground~$\ket{g}$ (Rydberg~$\ket{r}$) state~\cite{Homeier2023}. A coherent drive~$\Omega$ couples~$\ket{g}$ and~$\ket{r}$, and we choose the detuning in the facilitation-like regime,~$\Delta=V$~\cite{Marcuzzi2017}, with~$V$ the nearest-neighbor Rydberg-Rydberg interaction; here, we further neglect the van-der-Waals tails. This allows us to obtain a protection term~$\propto V$ that splits the energy manifold in vertices with even and odd number of Rydberg excitations, giving rise to a \Ztwo{}~Gauss' law measuring the local parity of excitations on the parent honeycomb lattice, see Fig.~\ref{figure-4}(c) and End Matter.


\textit{Discussion and Outlook.---}We have investigated the long-time mean-field dynamics of large-scale spin systems and uncovered a prethermal plateau with emergent \Ztwo{} gauge symmetry, whose breakdown follows a universal surface growth of nucleation regions characterized by non-linear growth exponents -- opening a path to explore KPZ dynamics in quantum simulators~\cite{Wei2022,Rosenberg2024}. These findings are crucial for understanding gauge violation errors and for developing universal frameworks for the thermalization of LGTs~\cite{Halimeh2025}. Our observations provide a guide to develop rigorous theories of slow thermalization in LGTs~\cite{Michailidis2018,Causer2024,Maric2026}. While ED confirms the plateau’s stability in small systems, existing numerical methods, including DTWA, cannot capture the quantum evolution, highlighting quantum simulators~\cite{Altman2021} -- such as Rydberg tweezer arrays, trapped ions, or superconducting qubits -- as promising platforms to experimentally probe these many-body dynamics and to benchmark our method for simulating large-scale gauge theories.

\textit{Acknowledgements.---}We thank Rahul Nandkishore, Alexander Baumgärtner, Adam Kaufman, Christian Groß, Joanna Lis, Hannes Pichler, Mu Qiao and Ruben Verresen for fruitful discussions.
LH and AMR acknowledge support by the Simons Collaboration on Ultra-Quantum Matter, which is a grant from the Simons Foundation (651440), and the NSF JILA-PFC PHY-2317149. This project has received funding from the European Research Council (ERC) under the European Union’s Horizon 2020 research and innovation programm (Grant Agreement no 948141) — ERC Starting Grant SimUcQuam. LH and FG were funded by the Deutsche Forschungsgemeinschaft (DFG, German Research Foundation) under Germany's Excellence Strategy -- EXC-2111 -- 390814868. AP acknowledges support by Trinity College Cambridge. HZ was funded by the Innovation Program for Quantum Science and Technology (No. 2024ZD0301800). J.C.H.~acknowledges funding by the Max Planck Society, the Deutsche Forschungsgemeinschaft (DFG, German Research Foundation) under Germany’s Excellence Strategy - EXC-2111 – 390814868, and the European Research Council (ERC) under the European Union’s Horizon Europe research and innovation program (Grant Agreement No.~101165667) -- ERC Starting Grant QuSiGauge. Views and opinions expressed are, however, those of the author(s) only and do not necessarily reflect those of the European Union or the European Research Council Executive Agency. Neither the European Union nor the granting authority can be held responsible for them.

%

\pagebreak
\clearpage
\newpage


\begin{center} 
\textbf{\large End matter}
\end{center}

\textit{Mean-field dynamics.---}
To derive the equations of motion at a mean-field level, we express the Hamiltonian~\eqref{eq:Hamiltonian} by classical spin operators~$\hat{\sigma}^\alpha_j \rightarrow \sigma^\alpha_j$. The dynamics is obtained from the Poisson bracket
\begin{align} \label{eq:DiffEq}
    \frac{d \vec{\sigma}_j}{dt} = \{ \vec{\sigma}_j, H \} = \vec{\sigma}_j \times \frac{\partial H}{\partial \vec{\sigma}_j},
\end{align}
where we use the angular momentum algebra of the spin operators.

To solve the set of differential equations, we apply a trotterization procedure in order to access large spin systems and long times. Here, the spins are rotated independently according to Eq.~\eqref{eq:DiffEq} for a small time~$\delta t = 0.01$, after which the effective magnetic fields are updated. We typically choose timesteps~$\delta t$ with $\delta t = 0.01 < 1/V_{\rm max}$, where~$V_{\rm max}=80$ is the largest protection strength considered in our study.

\begin{figure}[t!!]
\centering
\includegraphics[width=\linewidth]{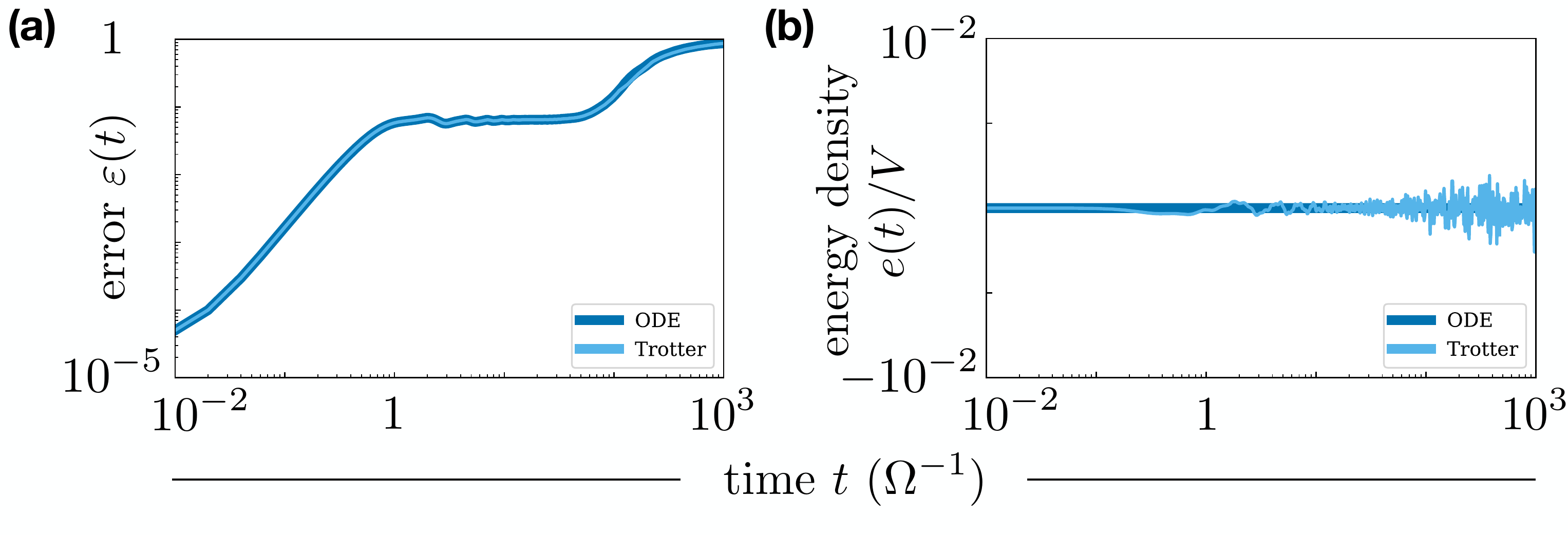}
\caption{\textbf{Numerical simulations.} We solve the set of differential equations~\eqref{eq:DiffEq} for a system with dynamical matter on a~$L=2\times 2$ plaquette lattice and periodic boundary conditions; the number of spins is~$N_{\rm spins} = 20$. We set the protection strength~$V/\Omega=10$ and $J = \Omega = 1$ and we consider a single trajectory in a random potential with~$\delta V=0.1V$ as well as a random initial configuration in the Gauss' law sector with~$G_j=+1$ for all~$j$. For the trotterization procedure, we choose the timestep to be~$\delta t = 0.01$. For the ODE solver, we use the MATLAB function \texttt{ode45}. \textbf{(a)} We plot the time-averaged Gauss' law error~$\varepsilon(t)$ solving the differential equations with an ODE solver and using the trotterization procedure. Only at late times ($t\Omega > 10^2 $) we observe slight deviations. \textbf{(b)} The equations-of-motion conserve energy~$E(t)=E(t=0)$ and hence we consider the energy density~$e(t) = E(t)/N_{\rm spins}$ as another observable to benchmark the validity of the trotterization procedure. For an initial gauge-invariant state, the energy  density is zero, $e(0)=0$. While the ODE solver exactly conserves energy, the trotterized time evolution shows small oscillations~$|e(t)| < 10^{-2} \cdot V$ at later times.}
\label{figureSI-trotter}
\end{figure}

In Fig.~\ref{figureSI-trotter} we benchmark the trotterization procedure against an ordinary differential equation (ODE) solver in small systems. First, we compare the time evolution of the Gauss' law error for a system of size~$L=2 \times 2$ with periodic boundary conditions, see Fig.~\ref{figureSI-trotter}(a), which shows excellent agreement up to exponentially long times, even on the level of four-point correlation functions, i.e., Gauss' law~$G_j$. Second, we compare the energy density~$e(t) = \langle \hat{H} \rangle/N_{\rm spins}$ -- a constant of motion -- for the two different methods, see Fig.~\ref{figureSI-trotter}(b). As expected, the ODE solver conserves energy up to all times. In the trotterization method we find small oscillations at late times, which are still controlled on the timescales we consider.

We emphasize that here we consider a single trajectory with fixed disorder realization and initial state. Comparing this dynamics to ED of the quantum model, which captures the full microscopic dynamics, can differ substantially from the mean-field dynamics on the level of a single trajectory. The thermalization of Gauss' law errors discussed in the main text occurs after averaging over many disorder realizations, where the mean-field
results quantitatively reproduce the ED behavior, see Fig.~\ref{figure-4}. 

\textit{Discrete Time Wigner Approximation.---}
In Fig.~\ref{figure-4}(a), we discuss the deviation of DTWA compared to mean-field and exact dynamics. Here, we further study larger systems of size~$L=20 \times 20$ as in Fig.~\ref{figure-1}(d). To this end, we average over $N_{\rm ens}=2000$ different realizations, where in each realization we choose (i) a random potential with~$\delta V = 0.1V$ and (ii) random gauge-invariant initial states with fluctuations in the $x$- and $y$-direction, so that the single-site quantum fluctuations~$\langle (\hat{\sigma}^x_\alpha)^2 \rangle$ = $\langle (\hat{\sigma}^y_\alpha)^2 \rangle = 1$ with $\alpha\in \{j,\ij\}$ at time~$t=0$ are correctly captured on average~\cite{Schachenmayer2015}. 

\begin{figure}[t!!]
\centering
\includegraphics[width=\linewidth]{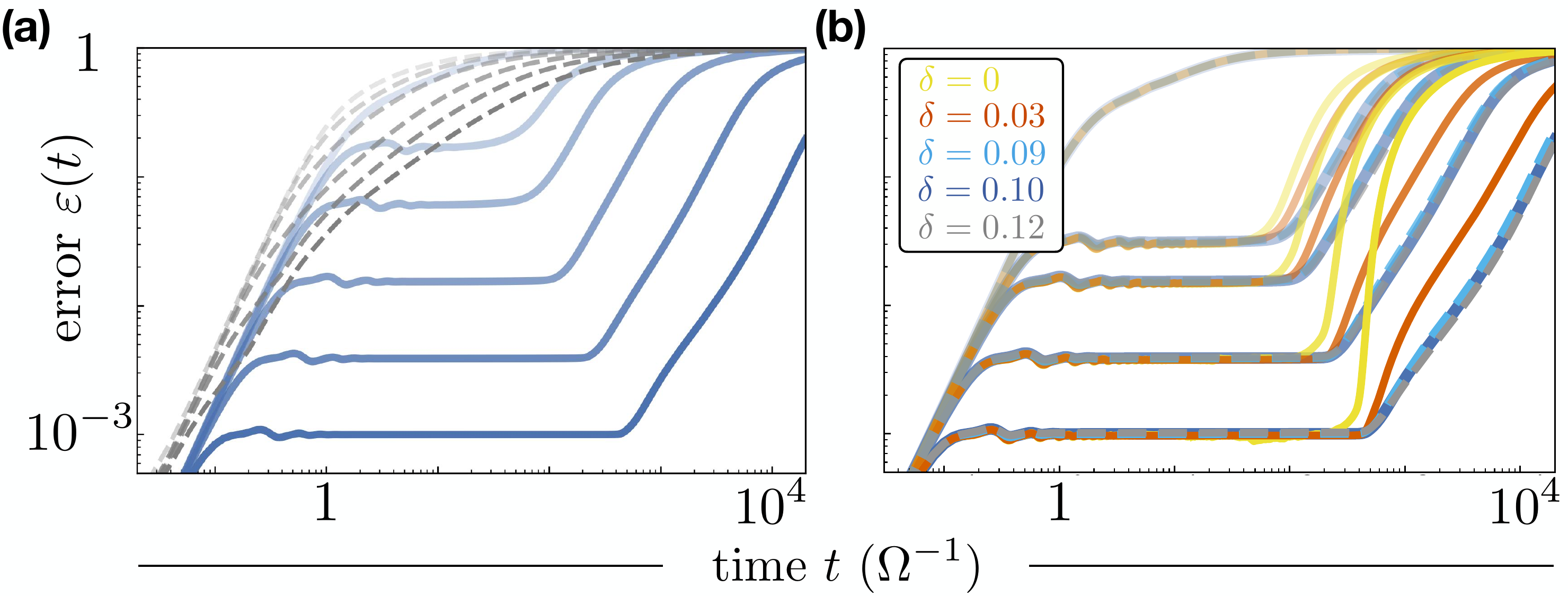}
\caption{\textbf{DTWA and random potentials.} \textbf{(a)} We consider the protocol from Fig.~\ref{figure-1}(d) and compare the mean-field dynamics to DTWA. Even at the largest protection strength~$V/\Omega=80$, DTWA does not capture a prethermal plateau. We highlight that ED does show a pronounced plateau, see Fig.~\ref{figure-4}. \textbf{(b)} We consider the protocol from Fig.~\ref{figure-1}(d) and study the effect of various disorder strengths with standard deviation~$\delta$, here for protection strength~$V/\Omega = 4,10,20,40,80$ (bright to dark). Without disorder, the late time thermalization shows a fast breakdown of gauge symmetry. At finite disorder, the thermalization dynamics is mostly unaffected by the value of~$\delta$. We highlight that the height of the plateau and the critical time, after which the plateau decays, is independent of the disorder strength.}
\label{figureSI-DTWA-Rand}
\end{figure}

As shown in Fig.~\ref{figureSI-DTWA-Rand}(a), the error~$\varepsilon(t)$ does not develop a plateau in the DTWA dynamics for any protection strength~$V/\Omega= 4,...,80$. Instead the system immediately reaches the thermal expectation value~$\varepsilon \rightarrow 1$. As discussed in Ref.~\cite{Muleady2023}, in the presence of symmetries the initial noise distribution in DTWA occupies multiple symmetry sectors and hence has limited performance. While in our case, the Hamiltonian does not have an explicit symmetry, the energetic constraints lead to an emergent local \Ztwo{}~symmetries. We attribute the failure of DTWA to capture the prethermal plateau to the occupation of gauge-symmetry violating sectors in the initial state leading to seeding of gauge defects in the system and hence a fast breakdown of Gauss' law.

\textit{Random potentials.---}
In Hamiltonian~\eqref{eq:Hamiltonian}, we introduce a vertex dependent disorder potential~$\delta_j$ with standard deviation~$\delta$. The disorder in the Hamiltonian fulfills two purposes: (i) It suppresses higher-order resonances between gauge-invariant and gauge-breaking states. (ii) The disorder in both the couplings and the initial states introduce spatial fluctuations in our system. For example, in a system without disorder, the line defect considered in Fig.~\ref{figure-3}, would lead to surface growth without any fluctuations in the height field~$h(X,t)$.

Here, we study the dependence of the thermalization dynamics on disorder strength~$\delta$. In Fig.~\ref{figureSI-DTWA-Rand}(b), we perform the same calculations as in Fig.~\ref{figure-1}(d), but with various disorder strengths~$\delta$. For the disorder-free case~$\delta=0$, we still find a prethermal plateau, however, with a slightly shorter lifetime and a fast decay. Upon increasing the disorder strength~$\delta$, the thermalization dynamics becomes independent of this parameter. 

\textit{Mean-field phase transition.---}
We consider the translationally-invariant model without dynamical matter ($\hat{\sigma}_{j}^z\equiv-1$ for all~$j$) starting from an initial state with all spins in~$\hat{\sigma}_{\ij}^z=-1$.
For this model, the Hamiltonian is given by
\begin{align}
    \hat{H} = V\sum_j \frac{1}{8} \left[\left( -1 + \sum_{i:\ij}\hat{\sigma}^z_{\ij} \right)^2 - 4 \right] + \frac{\Omega}{2} \sum_{\ij} \hat{\sigma}^x_{\ij},
\end{align}
with the Gauss' law,
\begin{align}
    \hat{G}_j = -(-1) \prod_{i: \ij} \hat{\sigma}^z_{\ij} = (\hat{\sigma}^z)^3,
\end{align}
where in the second step we have imposed the translational invariance of the problem.
The mean-field equation of motion effectively reduces to a Lipkin-Meshkov-Glick model~\cite{Lipkin1965},
\begin{align} \label{eq:LMG-model}
    \frac{d\vec{\sigma}}{dt} = \begin{bmatrix} \left( \frac{V}{2} - V\sigma^z \right)\sigma^y \\ -\left( \frac{V}{2} - V\sigma^z \right)\sigma^x - \frac{\Omega}{2}\sigma^z \\ \frac{\Omega}{2} \sigma^y \end{bmatrix}.
\end{align}
To simplify the set of differential equations, we use conservation of energy of the initial state to express
\begin{align} \label{eq:energy-conservation}
    \sigma^x = \frac{V}{\Omega}\left[ \sigma^z - (\sigma^z)^2 \right]
\end{align}
and the conservation of spin length to write
\begin{align} \label{eq:spin-length}
    (\sigma^y)^2 = 1 - (\sigma^x)^2 - (\sigma^z)^2.
\end{align}
This allows us to derive an equation of motion for the Gauss' law~$G = (\sigma^z)^3$ as
\begin{align} \label{eq:EOM-Gauss}
    \frac{dG}{dt} = 3 (\sigma^z)^2  \frac{d\sigma^z}{dt}.
\end{align}
Inserting Eqs.~\eqref{eq:LMG-model}-\eqref{eq:spin-length} into \eqref{eq:EOM-Gauss} yields a single differential equation for the Gauss' law dynamics,
\begin{align}
    \left(\frac{dG}{dt}\right)^2  + F(G) = 0
\end{align}
with the effective potential~$F(G)$ given by
\begin{align} \label{eq:effective-potential}
    F(G) = -\frac{9}{4}\Omega^2 \left[ G^{4/3} - \left( 1 + \frac{V^2}{\Omega^2}G^2 \right) - \frac{V^2}{\Omega^2}G^{8/3} + 2\frac{V^2}{\Omega^2} G^{7/3} \right],
\end{align}
which is plotted in Fig.~\ref{figure-2} for various~$V/\Omega$. The critical $(V/\Omega)_c$ is found by analyzing the roots of~$F(G)$.

\textit{Rydberg model without dynamical matter.---}
The model without dynamical matter can be realized in Rydberg arrays with atoms on the sites~$A,B$ of a Kagome lattice, i.e., the links of a honeycomb lattice. In particular, we map electric fields~$\sigma^z_{A} = +1$ on the interacting Rydberg state~$\ket{r_{A}}$ and~$\sigma^z_{A} = -1$ on an atomic ground state~$\ket{g_{A}}$; the former has density-density interactions and we set the nearest-neighbor interactions to have strength~$V$. 

We can replace the spin operators by~$\hat{\sigma}^z_{A}= 2\ket{r_A}\bra{r_A} -1 =: 2\hat{n}_A-1=-(-1)^{\hat{n}_A} $, where have used $\ket{r_A}\bra{r_A} + \ket{g_A}\bra{g_A} = 1$ and introduced a number operator~$\hat{n}_A=0,1$ to count the Rydberg excitation of an atom on site~$A$.
Inserting these identities in Hamiltonian~\eqref{eq:Rydberg_Ham} from the main text, yields
\begin{align} \label{eq:Rydberg_Ham_SI}
    \hat{H}^{\rm Ryd} = V\sum_{\langle A,B \rangle} \hat{n}_{A}\hat{n}_{B} - V\sum_A \hat{n}_{A} + \frac{\Omega}{2}\sum_A\hat{\sigma}^x_{A},
\end{align}
which is directly implementable in Rydberg arrays. The second term corresponds to the detuning~$\Delta=V$ of the drive~$\Omega$ from the bare atomic transition. Interactions between Rydberg atoms exhibit van-der-Waals tails~$\propto |A-B|^{-6}$, which we have neglected in our Hamiltonian. The thermalization dynamics is governed by the emergent \Ztwo{}~gauge structure and our construction of a metastable \Ztwo{}-invariant subspace remains intact under small diagonal energy shifts~\cite{Homeier2023}.

While it is convenient notation to describe the spins located on the Kagome lattice, we note that the Gauss' law is still defined on vertices of the honeycomb lattice. Assuming a static matter with $\hat{\sigma}^z_j\equiv 1$, we can write the Gauss' law for the Rydberg model as
\begin{align}
    \hat{G}^{\rm Ryd}_j = -\prod_{j:\ij}\hat{\sigma}^z_{\ij} = (-1)^{\sum_{j:\ij}\hat{n}_{\ij}},
\end{align}
showing that in the sector~$\hat{G}^{\rm Ryd}_j=+1$, the number of Rydberg excitations per vertex is even.

To obtain the energy landscape illustrated in Fig.~\ref{figure-4}(c), we write the first term in Hamiltonian~\eqref{eq:Rydberg_Ham_SI} as a sum of Hamiltonians~$\hat{H}_V = \sum_j \hat{H}_j$ for vertices~$j$ on the honeycomb lattice. Then each summand is comprised of all pairwise interactions of spins within this vertex and a detuning term. Hence, the detuning of a spin \textit{per vertex} is~$\Delta = V/2$, and each spin is associated with two vertices. On a honeycomb lattice, each vertex contains three atoms and hence a vertex with an odd number of Rydberg excitations ($\sum_{i: \ij} \hat{n}_{\langle i,j \rangle}= 1,3$) has a diagonal energy contribution of~$E_{\rm odd} = -V/2,3V/2$ and a vertex with an even number of Rydberg excitations ($\sum_{i: \ij} \hat{n}_{\langle i,j \rangle}= 0,2$) has an diagonal energy contribution~$E_{\rm even} = 0V$. In the limit of~$\Omega \ll V$, this separation into different energy sectors gives rise to the \Ztwo{}~gauge structure since the degenerate sector with even number of excitations per vertex is separated from the odd number sectors.


\begin{thebibliography}{74}
\bibitem [{Foo({\natexlab{a}})}]{Footnote_boundary_term}%
  \BibitemOpen
  \href@noop {} {\bibfield  {journal} {\bibinfo  {journal} {We include a
  pinning field on all sites in the top row to prevent spontaneous nuclation at
  the boundary.}\ } ({\natexlab{a}})}\BibitemShut {NoStop}%
\bibitem [{Foo({\natexlab{b}})}]{Footnote_euclidean}%
  \BibitemOpen
  \href@noop {} {\bibfield  {journal} {\bibinfo  {journal} {We have checked
  that a metric based on the Manhattan distance leads to qualitatively the same
  results.}\ } ({\natexlab{b}})}\BibitemShut {NoStop}%
\bibitem [{Foo({\natexlab{c}})}]{Footnote_three_body_term}%
  \BibitemOpen
  \href@noop {} {\bibfield  {journal} {\bibinfo  {journal} {Note that in the
  quantum system three-body dynamics is generated perturbatively. We have
  further checked that our results remain valid for $J = \Omega$.}\ }
  ({\natexlab{c}})}\BibitemShut {NoStop}%
\end{thebibliography}
\end{document}